\documentclass[aps,twocolumn]{revtex4}

\newcommand {\bea}{\begin{eqnarray}}
\newcommand {\eea}{\end{eqnarray}}
\newcommand {\be}{\begin{equation}}
\newcommand {\ee}{\end{equation}}

\usepackage{graphicx}

\begin{document}


\title{Low Energy Dynamics in Ultradegenerate QCD Matter}

\author{Thomas~Sch\"afer and Kai~Schwenzer}

\affiliation{Department of Physics, North Carolina State University,
Raleigh, NC 27695}

\begin{abstract}
 We study the low energy behavior of QCD Green functions in the 
limit that the baryon chemical potential is much larger than the QCD 
scale parameter $\Lambda_\mathrm{QCD}$. We show that there is a systematic 
low energy expansion in powers of $(\omega/m)^{1/3}$, where $\omega$ 
is the energy and $m$ is the screening scale. This expansion is valid
even if the effective quark-gluon coupling $g$ is not small. The 
expansion is purely perturbative in the magnetic regime $|\vec{k}|
\!\gg\! k_0$. If the external momenta and energies satisfy $|\vec{k}| 
\!\sim\! k_0$, planar, abelian ladder diagrams involving the full quark 
propagator have to be resummed but the corresponding Dyson-Schwinger 
equations are closed. 

\end{abstract}
\maketitle

  Landau's theory of Fermi liquids has been remarkably successful in 
many areas of physics \cite{Baym:1991}. Landau argued that 
although strongly interacting many-particle systems are very complicated 
in general, the situation greatly simplifies if the temperature is very 
low. In this limit the properties of the system can be understood in
terms of weakly interacting excitations called ``quasi-particles''.
The complicated dynamics of the underlying many-body systems 
can be summarized in terms of just a few parameters that 
characterize the quasi-particles and their scattering amplitudes. 

  In dense quark matter the presence of unscreened magnetic interactions 
invalidates the basic assumptions of Landau Fermi liquid theory \cite{Baym:uj,Vanderheyden:1996bw,Brown:2000eh,Holstein:1973}. 
Weak coupling calculations suggest that non-Fermi liquid corrections to 
the specific heat in the normal phase \cite{Ipp:2003cj,Schafer:2004zf} or 
the gap in the superfluid phase \cite{Son:1998uk} are substantial. 
However, weak coupling methods are restricted to densities that are much 
larger than the densities that can be achieved deep inside a neutron star.

 This raises the question whether there is a systematic low-energy 
expansion that can be used in much the same way that Landau theory is 
used in nuclear and condensed matter physics. There has been a lot of 
progress in this direction, mostly based on effective theories of dense QCD 
\cite{Hong:2000tn,Reuter:2004kk}. 
In this paper we show that QCD Green functions have low energy expansions in 
powers of $(\omega/m)^{1/3}$, where $\omega$ is the characteristic energy 
scale of the process and $m$ is the screening scale. This expansion is 
reliable as long as $\omega\ll m$ even if the gauge coupling $g$ is of order 
one. We provide a brief exposition of our main results. A detailed 
diagrammatic analysis will appear in a companion paper \cite{Schwenzer:2005xx}.

 Our starting point is an effective field theory (EFT) for fermions near 
the Fermi surface interacting with unscreened gauge fields. We follow 
Hong \cite{Hong:2000tn} and expand the fermionic part of the effective 
action in derivatives of the fields,
\bea
\label{l_hdet}
 \lefteqn{ {\cal L}_f    = \psi_{\pm\vec{v}}^\dagger  
  \left(  i Z_\parallel v_\pm\!\cdot\!D -
   Z_\perp \frac{D_\perp^2}{2\mu} + \delta\mu  \right)  
  \psi_{\pm\vec{v}} }&&  \\
&+& \!\frac{V_{\mathrm{\!ZS}}^\Gamma}{\mu^2} 
 (\psi_{\vec{v}}^\dagger\Gamma\psi_{\vec{v}}) 
 (\psi^\dagger_{\vec{v}} \Gamma\psi_{\vec{v}})\!+\!
 \frac{V_{\!\mathrm{BCS}}^\Gamma}{\mu^2} 
 (\psi_{\vec{v}}^\dagger\Gamma\psi_{\vec{v}}) 
 (\psi^\dagger_{\!-\vec{v}} \Gamma\psi_{\!-\vec{v}})\!+\!\cdots .
 \nonumber 
\eea
Here, $\psi_{\pm \vec{v}}$ describes particles and holes with momenta 
$p\!=\pm \!\mu\vec{v}\!+\!l$ and $v_\pm^\mu\!=\!(1,\pm \vec{v})$ labels 
the local Fermi velocity. We will write $l\!=\!l_0\!+\!l_{\|}\!+\!l_\perp$ 
where $l_{||}$ and $l_\perp$ are the components of $l$ parallel and
orthogonal to $\vec{v}$. We concentrate on two patches on opposite
sides of the Fermi surface as these are the only channels that lead
to kinematic enhancements. The effect of fluctuations above the cutoff 
$\Lambda$ is encoded in low energy constants. The leading coefficients
are the shift $\delta \mu$ of the Fermi energy, the Fermi velocity 
$v_F=|\vec{v}|$, the renormalization factors $Z_\parallel$ and $Z_\perp$, 
the gauge coupling $g$ at the scale $\Lambda$, and the four-fermion
couplings $V_{\mathrm{ZS}}^\Gamma$ and $V_{\mathrm{BCS}}^\Gamma$ in 
the forward (Zero Sound) and back-to-back (BCS) channels. 

 The dominant effect in the gluon sector is due to hard dense quark 
loops (HDLs) in gluonic correlation functions. These fluctuations involve 
small external energies $\omega$ but hard loop momenta $p\sim\mu$ in a 
narrow interval of width $\omega$. Hard dense loops are non-analytic in 
the gluon energy and momentum and cannot be represented as local operators. 
However, as long as hard fluctuations do not change the symmetries of the 
groundstate all other corrections are analytic. We can then write 
the gluonic lagrangian as ${\cal L}_g = {\cal L}_0 +{\cal L}_{HDL}+
\ldots$ where ${\cal L}_0$ is the free gluon lagrangian and 
${\cal L}_\mathrm{HDL}$ is the HDL generating functional 
\cite{Braaten:1989mz,Manuel:1995td}
\be
\label{S_hdl}
{\cal L}_{\mathrm{HDL}} = -\frac{m^2}{2}\sum_v \,G^a_{\mu \alpha}
  \frac{v^\alpha v^\beta}{(v\cdot D)^2}G^b_{\mu\beta} \, .
\ee
This term describes screening and damping of soft gluon modes due to 
particle-hole pairs on the entire Fermi surface. In perturbation theory 
the dynamical gluon mass is given by $m^2\!=\!N_f g^2\mu^2/(4\pi^2)$,
where $g$ is the QCD coupling constant. 


 We now analyze the dynamics of the theory governed by  
equ.~(\ref{l_hdet},\ref{S_hdl}). We first note that particle-hole
loops have already been integrated out and are represented by the
HDL term. The effective theory describes the interaction of particles
and holes with soft gluons which do not significantly change their 
velocity $\vec{v}$. Since electric fields are screened the interaction 
at low energies is dominated by the exchange of magnetic gluons. 
Magnetic gluons are weakly damped in the kinematic regime $|k_0|\!
\ll\!|\vec{k}|$. In this regime 
\be
D^{(m)}_{ij}(k) = \frac{\delta_{ij}-\hat{k}_i\hat{k}_j}{k_0^2-\vec{k}^2+
i\frac{\pi}{2}m^2 |\frac{k_0}{\vec{k}}|} \, .
\ee
and we observe that the propagator becomes large if
\be 
\label{ld_kin} |\vec{k}| \sim (m^2 |k_0|)^{1/3} \gg |k_0| \, .
\ee
This implies that the gluon is very far off its energy shell and not 
a propagating state. We will compute a general diagram by picking up 
the pole in the quark propagator, and integrate over the cut in the 
gluon propagator using the kinematics dictated by equ.~(\ref{ld_kin}). 
In order for a quark to absorb the large momentum carried by a gluon 
and stay close to the Fermi surface this momentum has to be transverse 
to the momentum of the quark. This means that the term $k_\perp^2/(2\mu)$ 
in the quark propagator is relevant and has to be kept at leading order. 
Equation (\ref{ld_kin}) shows that $k_\perp^2/(2\mu)\!\gg\!k_0$ as
$k_0\!\to\! 0$. This means that the pole of the quark propagator is 
governed by the condition $k_{||}\!\sim\! k_\perp^2/(2\mu)$. We 
conclude that quark and gluon momenta scale with respect to an 
external energy scale $\omega$ as
\be
\label{ld_reg} 
 k_0 \sim \omega \, ,\quad 
 k_{||}  \sim m^\frac{4}{3} \omega^\frac{2}{3}/\mu \, , \quad 
 k_\perp \sim m^\frac{2}{3} \omega^\frac{1}{3} \, .
\ee
We will refer to the regime in which all momenta, including external 
ones, satisfy the scaling relation (\ref{ld_reg}) as the magnetic 
regime. A similar regime was identified in the context of gauge theories
of condensed matter systems \cite{Polchinski:ii}. The 
scaling relations (\ref{ld_reg}) are the basis of the low energy 
expansion in ultradegenerate matter. 

In the low energy regime propagators and vertices can be simplified
even further. The quark and gluon propagators are 
\bea 
\label{quarkprop}
 &&  S_{\pm \vec{v}}^{\alpha\beta}(p) 
   = \frac{i \delta_{\alpha\beta}}
         {Z_\parallel \left( p_0\mp v_F p_{||} \right) 
         - Z_\perp \frac{p_\perp^2}{2\mu}
         +i\epsilon {\mathrm{sgn}}(p_0)} \, , \;\;\; \\ 
\label{gluonprop}
 &&  v_+^\mu v_\pm^\nu D^{(m)}_{\mu \nu}(k)  = \mp \frac{iv_F^2}
       {k_\perp^2+i\frac{\pi}{2}m^2\frac{\left|k_0\right|}{k_\perp}} \, , 
\eea
and the quark gluon vertex is $gZ_\parallel v_i(\lambda^a/2)$. Higher order 
corrections can be found by expanding the quark and gluon propagators 
as well as the HDL vertices in powers of the small parameter $\epsilon\!
\equiv\!\omega/m$ \cite{Schwenzer:2005xx}. We observe that the transverse 
projector in the gluon propagator simplifies because $k_\perp\!\gg\!k_{||}$. 
We also note that in the magnetic regime the factor $p_0$ in the quark 
propagator can be dropped since $p_0\!\ll\!p_{||}$.

Using these expressions we can show that the power of $\epsilon$
associated with a Feynman diagram always increases with the number 
of loops and the number of higher-order vertices. One way to see 
this is to rescale the fields in the effective lagrangian so that 
the kinetic terms are scale invariant under the transformation 
$(x_0,x_{||},x_\perp)\to (\epsilon^{-1}x_0,\epsilon^{-2/3} x_{||},
\epsilon^{-1/3}x_{\perp})$. The scaling behavior of the fields is 
$\psi\to\epsilon^{5/6}\psi$ and $A_i  \to \epsilon^{5/6} A_i$. We find 
now that the scaling dimension of all interaction terms is positive. 
The quark gluon vertex scales as $\epsilon^{1/6}$, the HDL three gluon 
vertex scales as $\epsilon^{1/2}$, and both the quark-two-gluon vertex
induced by the $D_\perp^2$ term in equ.~(\ref{l_hdet}) and the four
gluon vertex scale as $\epsilon$. Since higher order diagrams involve at 
least one pair of quark gluon vertices the expansion involves positive 
powers of $\epsilon^{1/3}$ and the magnetic regime is completely 
perturbative.

\begin{figure}
\includegraphics[width=3.25cm]{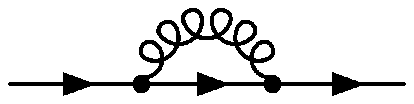}\hspace{-0.75cm}
\includegraphics[width=3.25cm]{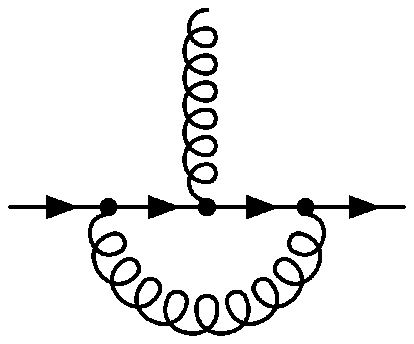}\hspace{-0.75cm}
\includegraphics[width=3.25cm]{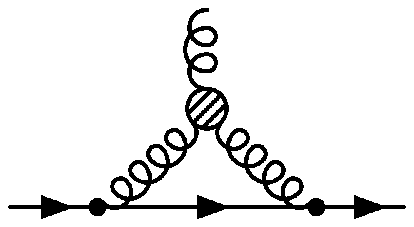}
\caption{One-loop contributions to the quark self energy 
and the quark-gluon vertex. In the magnetic regime  
the graphs scale as $\omega\log(\omega)$, $\omega^{1/3}$
and $\omega^{2/3}$, respectively. }
\label{fig_mag}
\end{figure}

 As a simple example we consider the fermion self energy. 
The one-loop diagram is 
\bea
\Sigma(p) &=& g^2 C_F\int \frac{dk_0}{2\pi}
 \int \frac{dk_\perp^2}{(2\pi)^2} \ 
    \frac{k_\perp}{k_\perp^3+i\frac{\pi}{2} m^2 |k_0|}  \nonumber \\
 &\times&\!\!\int \frac{dk_{||}}{2\pi} \  \frac{\Theta(p_0+k_0)}
   {(k_{||}+p_{||})
  +\frac{Z_\perp (k_\perp+p_\perp)^2}{2Z_\parallel v_F\mu}+i\epsilon} \, ,  
\eea
with $C_F\!=\!(N_c^2\!-\!1)/(2N_c)$. This expression shows a number of 
interesting features. First we observe that the longitudinal and transverse 
momentum integrations factorize. The longitudinal momentum integral is 
performed by picking up the pole in the quark propagator. The result is 
independent of the external momenta and only depends on the external energy. 
The transverse momentum integral is logarithmically divergent. 
We find \cite{Vanderheyden:1996bw,Brown:2000eh}
\be 
\label{sig_m}
 \Sigma(p) = \frac{g^2}{9\pi^2} 
    \left( \omega \log\left(\frac{\Lambda_\Sigma}{|\omega|} \right) 
       \!+\! \omega \!+\! i\frac{\pi}{2}|\omega| \right)
       \!+\! O\left(\epsilon^\frac{5}{3}\right) ,
\ee
where $\omega \!\equiv\! p_0$. We have absorbed the logarithmic cutoff 
dependence into the low energy constant $Z_\parallel$. In general this 
result depends on two unknown parameters $g$ and $\Lambda_\Sigma$ where 
$\Lambda_\Sigma\!=\!\Lambda\exp(9\pi^2(Z_{||}\!-\!1)/(g^2Z_{||} v_F))$ 
and $\Lambda =2\Lambda_\perp^3/(\pi m^2)$ is related to the transverse
momentum cutoff. If the coupling is small, the scale is determined 
by the exchange of electric gluons and we find $\Lambda_{\Sigma} \!=\! 
2^{5/2}m/\pi$. We observe that the self energy correction is large and 
has to be included in the propagator whenever its energy dependence is 
relevant. We showed previously that rainbow diagrams do not give corrections 
of the form $g^{2n}\omega \log(\omega)^n$ \cite{Schafer:2004zf}. 
Equ.~(\ref{sig_m}) shows that higher order corrections are suppressed 
by powers of $\epsilon^{2/3}$ \cite{Gerhold:2005uu}. 

\begin{figure}
\includegraphics[width=3.5cm]{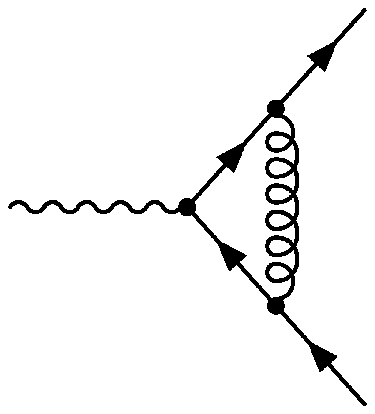}
\includegraphics[width=3.5cm]{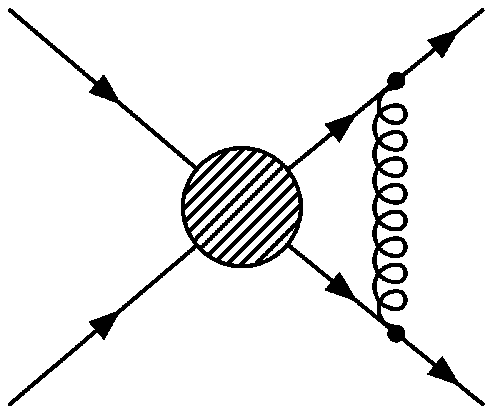}
\caption{One-loop correction to the vertex of an external 
current and the BCS interaction. Both diagrams are kinematically
enhanced and scale as $\log(\omega)$ and $\log^2(\omega)$, respectively. }
\label{fig_enh}
\end{figure}

 The scaling arguments apply to arbitrary Green functions in the 
magnetic regime, see Fig.~\ref{fig_mag}. Exceptions occur if the external 
fields have small momenta of the order of the external energy scale. This 
situation can occur in quark-quark scattering amplitudes or in 
vertex functions for external currents like the weak interaction 
\cite{Schafer:2004jp}. Consider the one-loop vertex correction 
for a color singlet vertex $\Gamma_\mu=eZ_\parallel v_\mu$. In the 
magnetic regime the graph scales like $\epsilon^{1/3}$ . This is 
confirmed by an explicit calculation \cite{Schwenzer:2005xx}. In the 
time-like regime $p_0\!-\!q_0> |\vec{p}\!-\!\vec{q}|$ 
\bea
\label{vertex_2}
\Gamma^\mu \left(p,q\right) = eg^2 C_F \hat{v}_+^\mu
      \int \frac{dk_0}{2\pi}\int \frac{d^2k_\perp}{(2\pi)^2} 
      v_+^\rho v_+^\sigma D^{(m)}_{\rho \sigma}(k) \nonumber \\
\times \int \frac{dk_{||}}{2\pi} Z_\parallel^2
  S_{\vec{v}}\left(k+p\right) S_{\vec{v}}\left(k+q\right), \qquad
\eea
where $p$, $q$ are the momenta of the external quarks. The important point 
is that if we combine the fermionic propagators using Fermi's trick in 
order to resolve the pole in the longitudinal momentum integration, 
the large components $k_{||}$ and $k_\perp^2/ (2\mu)$ of the 
propagators cancel and the result becomes sensitive to the small scales 
$p$, $q$
\[
 S_{\vec{v}}\left(k\!+\!p\right) S_{\vec{v}}\left(k\!+\!q\right)
   \!=\!\frac{S_{\vec{v}}\left(k+p\right)-S_{\vec{v}}\left(k+q\right)}
             {S_{\vec{v}}^{-1}(p)\!-\!S_{\vec{v}}^{-1}(q)
  \!-\!\frac{2 \vec{k}_{\perp}\!\cdot(\vec{p}_\perp +\vec{q}_\perp)}{\mu}} , 
\]
where constant factors are suppressed.
As a consequence the result is enhanced by a factor $1/\epsilon^{1/3}$. 
This enhancement is analogous to the one occurring in the HDL case.
In the limit $p_0\!-\!q_0\!\to\!0$ the $k_{||}$ integral gives a factor
$\delta(p_0-k_0)$ and the vertex correction is \cite{Brown:2000eh}
\be 
 \Gamma^\mu (p,q) = \frac{eg^2}{9\pi^2} \hat{v}_+^\mu 
 \log\left(\frac{\Lambda_{ZS}}{\left| \omega \right|}\right)  \, ,
\ee
where $\omega\!=\!(p_0\!+\!q_0)/2$. The logarithmic divergence was removed 
by adding the contribution from the four-fermion vertex in the zero 
sound channel. If the coupling is weak the scale inside the logarithm 
is again determined by electric gluon exchange. We find that in this 
case the scale is equal to the scale in the quark self energy. 

 The cancellation that occurs in the one-loop diagram repeats itself 
at any loop order if additional gluon ladders are added. This implies 
that ladder diagrams have to be summed. We also note that quark propagators 
in the ladders are sensitive to the small scale $\omega$ and thereby 
the full fermion self energy has to be included. A detailed analysis 
shows that all other corrections like 
crossed ladders, vertex corrections, interconnections between the gluon 
ladders, etc. introduce extra transverse momenta and follow the scaling 
relations in the magnetic regime \cite{Schwenzer:2005xx}.

 The same phenomenon occurs in the quark-quark scattering amplitude. 
Consider a one-loop correction to the scattering amplitude in 
the BCS-channel, see Fig.~\ref{fig_enh} 
\bea
\label{bcs_1l}
\delta\Gamma_{\mathrm{BCS}}^{\Gamma} (p,q) 
   = g^2C_{\Gamma}  \int \frac{dk_0}{2\pi}
         \int \frac{d^2k_\perp}{(2\pi)^2} \,
          v_+^\rho v_-^\sigma D^{(m)}_{\rho \sigma}(k-q)    
  \nonumber \\
  \times  \Gamma_{\mathrm{BCS}}^{\Gamma} (p,k) 
     \int \frac{dk_{||}}{2\pi} Z_\parallel^2 
   S_{\vec{v}}\left(k+p\right) S_{-\vec{v}}\left(-k+p\right),\quad
\eea
Here $p \pm q$ are the momenta of the outgoing fermions and
$C_\Gamma$ is a color factor. In the color anti-symmetric channel 
$C_\Gamma \!=\!-(N_c\!+\!1)/(2N_c)$. Combining the quark propagators 
gives
\[
 S_{\vec{v}}\left(k\!+\!p\right) S_{-\vec{v}}\left(-k\!+\!p\right)
  \!=\! \frac{S_{\vec{v}}\left(k+p\right)-S_{-\vec{v}}\left(-k+p\right)}
   {2k_0\!+\!S_{\vec{v}}^{-\!1}(p)\!-\!S_{-\vec{v}}^{-\!1}(p)
  \!-\! \frac{2 \vec{k}_{\perp}\!\cdot\vec{p}_{\perp}}{\mu}} . \nonumber
\]
This denominator becomes sensitive to the energy $k_0$ when the external 
quarks are on-shell and the transverse momentum of the pair vanishes.
We get 
\be 
\delta\Gamma_{\mathrm{BCS}}^{\bar 3} (p,q)=
  \frac{g^2}{18\pi^2}
   \int_{\omega} \frac{dk_0}{k_0}
  \Gamma_{\mathrm{BCS}}^{\bar 3} (p,k)
   \log\left( \frac{\Lambda_{BCS}}{|k_0-q_0|} \right) \ .
\ee
which again depends to leading order on $\omega\!\equiv\!p_0$.
The cutoff dependence was removed by adding the one-loop graph with 
the BCS four-fermion operator. We can compute the scale in 
the weak coupling limit. In the quark self energy and vertex correction 
the logarithmic UV divergence cancels between electric and 
magnetic gluon exchanges and the scale is determined by transverse 
momenta $k_\perp\!\sim\! m$. In the BCS channel the two terms add 
and the logarithm is sensitive to larger scales $k_\perp\!\sim\!\mu$. 
Matching to the full tree level scattering amplitude at this scale 
gives $\Lambda_{\mathrm{BCS}}\!=\!2^{11/2}\mu^6/(\pi m^5)$ as shown 
in \cite{Schafer:1999jg}. 

\begin{table}[t]
\begin{center}
\begin{tabular}{|c|c|c|c|} \hline
   $g(\Lambda)$ 
 & $m$ 
 & $\delta\mu$
 &$v_F$ \\ \hline
   $g(\mu)$ 
 & $\, \frac{N_f^{1/2}g}{2\pi}\mu \,$ 
 & $\; \frac{g^2}{3\pi^2}\mu \; $
 & $1-\frac{g^2}{9\pi^2}\log \left( \frac{2^{5/2} \mathrm{e}^{3/2} m}
       {\pi \Lambda} \right)$\\ \hline
\end{tabular}

\vspace*{0.3cm}
\begin{tabular}{|c|c|c|c|} \hline
 $Z_\parallel$
 & $Z_\perp$
  &  $V_{\mathrm{ZS}}^1$     
   &  $V_{\mathrm{BCS}}^{\bar 3}$ \\ \hline
  $1+\frac{g^2}{9\pi^2}\!\log \! 
      \left( \frac{2^{5/2} m}{\pi \Lambda} \! \right)$ 
  & $1-\frac{g^2}{6\pi^2}$
   & $\frac{2g^2}{9}\log \! 
      \left( \! \frac{2^{5/2} m}{\pi \Lambda} \! \right)$ 
   & $\frac{g^2}{9} \!\log \! \left( \!  
   \frac{2^{11/2} \mu^6}{\pi m^5 \Lambda} \! \right)$\\
\hline
\end{tabular}
\vspace{0.1cm}
\caption{Leading order expressions for the low energy constants 
appearing in the effective Lagrangian for magnetic modes 
equ.~(\ref{l_hdet}) in the weak coupling limit.}
\label{tab_cons}
\end{center}
\end{table}

\begin{figure}
\includegraphics[width=9.0cm]{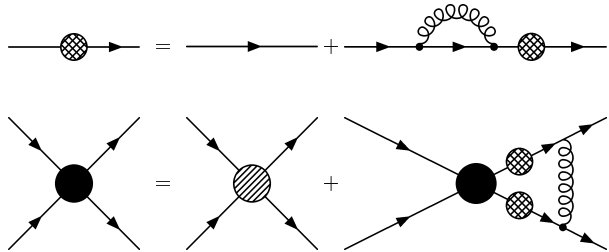}
\caption{The complete dynamics contributing to the Dyson-Schwinger 
equation for the BCS kernel in the low energy limit. The dark blob 
represents the full scattering amplitude and the shaded blob the full 
propagator. An analogous but independent equation is obtained for 
the forward channel.}
\label{fig_bs}
\end{figure}

 The double logarithmic structure of the quark-quark scattering 
amplitude in QCD was discovered by Son \cite{Son:1998uk}. 
Again we find that all planar, abelian ladder diagrams contribute 
at the same order but other corrections are perturbative and follow 
the scaling rules in the magnetic regime. The ladder sum can be 
determined by solving a Bethe-Salpeter equation, see Fig.~\ref{fig_bs}.
Since the quark propagators are sensitive to the low energy 
scale the quark self energy has to be included to all orders. 
We saw, however, that the one-loop result for the quark self 
energy is exact in the low energy limit. The situation simplifies 
further if the coupling is small. In this case we find that the 
low energy scale in  the BCS channel is $\omega \sim \Delta \sim 
\mu \exp(-1/g)$. This means that the quark self energy correction 
never becomes large and can be treated perturbatively, too. 

The analysis in the forward scattering channel is identical 
to the case of an external current. Green functions with more 
than four external quark lines do not have kinematic enhancements
except for those that occur in four-particle reducible graphs. 
The reason is that whenever a quark line is connected to a
given graph via a gluon an extra transverse momentum is introduced
which leads to suppression factors at low energy consistent with 
the scaling rules. 

 In summary, we have shown that QCD Green functions in high
density QCD have a systematic low energy expansion in powers
of $(\omega/m)^{1/3}$. The results are valid in any phase in
which the transverse components of the gauge field are dynamically
damped, ${\rm Im} \Pi(k_0, \vec{k})\!\sim\! m^2 k_0/|\vec{k}|$.
The damping scale $m$ determines the asymptotic behavior of all
gluonic correlation functions and can be used to define $m$
non-perturbatively. 
We note, in particular, that the low energy
interaction between gluons is weak and no magnetic screening mass
is generated in the normal phase. Our results apply likewise to ordinary condensed matter systems described by a strong effective gauge interaction. \cite{Polchinski:ii}.


 The low energy expansion is directly applicable to the normal phase 
of dense quark matter, and to modes that remain ungapped below the 
(largest) critical temperature. Such ungapped modes play a vital role for a potential detection of quark matter in compact stars. The effective theory can be used to 
study transport properties, neutrino emissivities \cite{Schwenzer:2006hk}, etc. 
Under conditions appropriate to neutron star 
cores we have $\left(T/\mu\right)^{1/3} \!\sim\!1/10$ and the 
expansion is expected to converge well. This provides robust QCD results on the generic temperature dependence of low energy observables which due to the huge ratio $\mu/T$ are not impaired by the ignorance of low energy constants of order one. 

 For $T<T_c\sim\Delta$ dense quark matter becomes superfluid, and the 
Fermi liquid description breaks down completely. For energies less 
than the gap, the non-Fermi liquid EFT has to be matched against an 
effective theory of the superfluid phase \cite{Casalbuoni:1999wu}.
The approximation scheme described here can be
used to compute the parameters of the superfluid EFT in an effective expansion controlled by parameters of magnitude $(\Delta/m)^{1/3}$ and $1/\log(m/\Delta)$. 
We note, in particular, that $g$ decreases only logarithmically with $\mu$, while the low energy 
expansion contains inverse powers of $\mu$. Therefore, this expansion represents a significant improvement over the mere use of perturbation theory in $g$ and extends the range of validity of perturbative QCD to considerably lower densities.

Acknowledgments: This work was supported in part by
US DOE grant DE-FG-88ER40388. 



\begin{thebibliography}{20}

\bibitem{Baym:1991}
G.~Baym and C~J.~Pethick, 
Landau Fermi Liquid Theory: Concepts and Applications,
J.~Wiley and Sons, New York, 1991;
G.~Baym and S.~A.~Chin,
Nucl.\ Phys.\ A {\bf 262}, 527 (1976).

\bibitem{Baym:uj}
G.~Baym, H.~Monien, C.~J.~Pethick and D.~G.~Ravenhall,
Phys.\ Rev.\ Lett.\  {\bf 64}, 1867 (1990).

\bibitem{Vanderheyden:1996bw}
B.~Vanderheyden and J.~Y.~Ollitrault,
Phys.\ Rev.\ D {\bf 56}, 5108 (1997);
C.~Manuel,
Phys.\ Rev.\ D {\bf 62}, 076009 (2000).

\bibitem{Brown:2000eh}
W.~E.~Brown, J.~T.~Liu and H.~Ren,
Phys.\ Rev.\ D {\bf 62}, 054013 (2000).

\bibitem{Holstein:1973}
T.~Holstein, A.~E.~Norton, P.~Pincus;
Phys.\ Rev.\ {\bf B8}, 2649 (1973).
M.~Yu.~Reizer, 
Phys.\ Rev.\ {\bf B 40}, 11571 (1989);
S.~Chakravarty, R.~E.~Norton and O.~F.~Syljuasen,
Phys.\ Rev.\ Lett.\  {\bf 75}, 1423 (1995).

\bibitem{Ipp:2003cj}
A.~Ipp, A.~Gerhold and A.~Rebhan,
Phys.\ Rev.\ D {\bf 69}, 011901 (2004).

\bibitem{Schafer:2004zf}
T.~Sch{\"a}fer and K.~Schwenzer,
Phys.\ Rev.\ D {\bf 70}, 054007 (2004).

\bibitem{Son:1998uk}
D.~T.~Son,
Phys.\ Rev.\ D {\bf 59}, 094019 (1999);
W.~E.~Brown, J.~T.~Liu and H.~C.~Ren,
Phys.\ Rev.\ D {\bf 61}, 114012 (2000).

\bibitem{Hong:2000tn}
D.~K.~Hong,
Phys.\ Lett.\ B {\bf 473}, 118 (2000);
Nucl.\ Phys.\ B {\bf 582}, 451 (2000);
G.~Nardulli,
Riv.\ Nuovo Cim.\  {\bf 25N3}, 1 (2002);
T.~Sch{\"a}fer,
Nucl.\ Phys.\ A {\bf 728}, 251 (2003).

\bibitem{Reuter:2004kk}
P.~T.~Reuter, Q.~Wang and D.~H.~Rischke,
Phys.\ Rev.\ D {\bf 70}, 114029 (2004).

\bibitem{Schwenzer:2005xx}
T.~Sch{\"a}fer and K.~Schwenzer, 
in preparation. 

\bibitem{Braaten:1989mz}
E.~Braaten and R.~D.~Pisarski,
Nucl.\ Phys.\ B {\bf 337}, 569 (1990);
Phys.\ Rev.\ D {\bf 45}, 1827 (1992).

\bibitem{Manuel:1995td}
C.~Manuel,
Phys.\ Rev.\ D {\bf 53}, 5866 (1996).

\bibitem{Polchinski:ii}
J.~Polchinski,
Nucl.\ Phys.\ B {\bf 422}, 617 (1994);
C.~Nayak and F.~Wilczek,
Nucl.\ Phys.\ B {\bf 430}, 534 (1994).

\bibitem{Gerhold:2005uu}
  A.~Gerhold and A.~Rebhan,
  Phys.\ Rev.\ D {\bf 71} (2005) 085010.

\bibitem{Schafer:2004jp}
T.~Sch{\"a}fer and K.~Schwenzer,
Phys.\ Rev.\ D {\bf 70}, 114037 (2004).

\bibitem{Schafer:1999jg}
T.~Sch{\"a}fer and F.~Wilczek,
Phys.\ Rev.\ D {\bf 60}, 114033 (1999);
R.~D.~Pisarski and D.~H.~Rischke,
Phys.\ Rev.\ D {\bf 61}, 074017 (2000);
D.~K.~Hong, V.~A.~Miransky, I.~A.~Shovkovy and L.~C.~R.~Wijewardhana,
Phys.\ Rev.\ D {\bf 61}, 056001 (2000).

\bibitem{Schwenzer:2006hk}
  K.~Schwenzer,
  PoS {\bf JHW2005} (2006) 030.

\bibitem{Casalbuoni:1999wu}
R.~Casalbuoni and R.~Gatto,
Phys.\ Lett.\ B {\bf 464}, 111 (1999).

\end{thebibliography}
\end{document}